*Hypothesis Article*

# The Venusian Lower Atmosphere Haze as a Depot for Desiccated Microbial Life: A Proposed Life Cycle for Persistence of the Venusian Aerial Biosphere

Sara Seager,[1–3] Janusz J. Petkowski,[1] Peter Gao,[4] William Bains,[1] Noelle C. Bryan,[1] Sukrit Ranjan,[1] and Jane Greaves[5,6]

## Abstract

We revisit the hypothesis that there is life in the Venusian clouds to propose a life cycle that resolves the conundrum of how life can persist aloft for hundreds of millions to billions of years. Most discussions of an aerial biosphere in the Venus atmosphere temperate layers never address whether the life—small microbial-type particles—is free floating or confined to the liquid environment inside cloud droplets. We argue that life must reside inside liquid droplets such that it will be protected from a fatal net loss of liquid to the atmosphere, an unavoidable problem for any free-floating microbial life forms. However, the droplet habitat poses a lifetime limitation: Droplets inexorably grow (over a few months) to large enough sizes that are forced by gravity to settle downward to hotter, uninhabitable layers of the Venusian atmosphere. (Droplet fragmentation—which would reduce particle size—does not occur in Venusian atmosphere conditions.) We propose for the first time that the only way life can survive indefinitely is with a life cycle that involves microbial life drying out as liquid droplets evaporate during settling, with the small desiccated ''spores'' halting at, and partially populating, the Venus atmosphere stagnant lower haze layer (33–48 km altitude). We, thus, call the Venusian lower haze layer a ''depot'' for desiccated microbial life. The spores eventually return to the cloud layer by upward diffusion caused by mixing induced by gravity waves, act as cloud condensation nuclei, and rehydrate for a continued life cycle. We also review the challenges for life in the extremely harsh conditions of the Venusian atmosphere, refuting the notion that the ''habitable'' cloud layer has an analogy in any terrestrial environment. Key Words: Venus— Clouds—Life—Habitability—Sulfuric Acid—Life Cycle—Aerial Biosphere. Astrobiology 21, xxx–xxx.

## 1. Introduction and Overview

L<small>IFE ON</small> V<small>ENUS</small> has been a topic of speculation for more than half a century, with published papers ranging from science-fiction-like to invalid conjecture to legitimate hypothesis (Morowitz and Sagan, 1967; Grinspoon, 1997; Cockell, 1999; Schulze-Makuch and Irwin, 2002, 2006; Schulze-Makuch *et al.*, 2004; Grinspoon and Bullock, 2007; Limaye *et al.*, 2018). Today, only Venus' atmospheric cloud layers (a large region spanning from 48 to 60 km altitude) have seemingly habitable conditions—the surface (at 735 K) is too hot for any plausible solvent and for most organic covalent chemistry. How the clouds could become inhabited is not known. In principle, life could arise in the clouds independent from the ground (Woese, 1979; Dobson *et al.*, 2000) with material from meteoritic input (Sleep, 2018a, 2018b) from the asteroid belt (including Ceres, and even from Mars). Life may even have been directly seeded by impacts from Earth ejecta (Melosh, 1988; Reyes-Ruiz *et al.*, 2012; Beech *et al.*, 2018). A more commonly agreed on, and perhaps more conceivable scenario, is that life originated on the surface, as it most likely did on Earth, and migrated into the

Departments of [1]Earth, Atmospheric and Planetary Sciences, [2]Physics, [3]Aeronautics and Astronautics, Massachusetts Institute of Technology, Cambridge, Massachusetts, USA.
[4]Department of Astronomy, University of California at Berkeley, California, USA.
[5]School of Physics and Astronomy, Cardiff University, Cardiff, United Kingdom.
[6]Institute of Astronomy, Cambridge University, Cambridge, United Kingdom.

clouds. Recent modeling by Way *et al.* (2016) suggests the existence of habitable surface and oceans as late as ∼700 Mya. Consideration of life on Venus is extensively summarized in a recent paper (Limaye *et al.*, 2018).

Almost all previous work on life in Venusian clouds does not specify what exactly ''life in the clouds'' means (for one exception see Schulze-Makuch *et al.*, 2004). Does microscopic life reside inside cloud liquid droplets? Or is microscopic life free floating in the air between cloud droplets? Earth has an aerial biosphere created by microbial life that regularly migrates to clouds from the ground (Vaïtilingom *et al.*, 2012; Amato *et al.*, 2017). Studies of Earth's aerial biosphere shows that microbes mostly reside inside cloud droplets but some are free-floating in the atmosphere (Section 4.1). Some microbes (both inside and outside droplets) are found to be metabolically active, even though there is no evidence, as of yet, of cell division (Amato *et al.*, 2019). Microbial life cannot reside in Earth's atmosphere indefinitely, mostly because of lack of continuous cloud cover, and after days to weeks microbes are deposited back down on Earth's surface (Burrows *et al.*, 2009; Bryan *et al.*, 2019). Earth's aerial biosphere is intimately connected to the habitable surface of the planet.

We argue that life, if it exists in Venus' atmosphere, must reside inside cloud liquid droplets for the majority of its life cycle (Section 2.1). Life engulfed by cloud droplets will be protected from a fatal net loss of liquid to the atmosphere, an unavoidable problem for any free-floating microbial lifeforms. But a droplet habitat implies a lifetime limitation. As liquid droplets coalesce and grow, they eventually reach a size that, due to gravity, settles out of the temperate layers of the atmosphere at an appreciable rate. Over time, the population of inhabited droplets should therefore decline to zero. (Droplet fragmentation—which would reduce particle size—does not occur in Venusian atmosphere conditions [Section 3.4].) The conundrum is that there is no way for liquid droplets—and hence life inside of them—to persist indefinitely in Venus' temperate atmosphere layers.

In this article, we describe this problem, and a solution to it. We hypothesize how life escapes being ''rained out'' down to inhospitably hot atmosphere layers by cycling between small, desiccated spores and larger, metabolically active, droplet-inhabiting cells. Venusian life escapes settling to the surface by forming a resistant, spore-like form that survives the evaporation of the inevitable downward droplet flow to atmosphere layers of high temperature. The desiccated spores become suspended in the Venus atmosphere lower haze layer, which we thus call a ''depot'' for desiccated microbial life. Because the dynamics of the relatively stagnant lower haze layer are not well known, the main uncertainty in our life cycle hypothesis is how the spores are transported back up into the clouds again. The spores most likely travel upward by vertical mixing induced by gravity waves, and once in the cloud layer they form the nucleus of a new droplet. The depot is ''leaky,'' that is spores will also vertically mix downward to atmosphere layers with fatally high temperatures. Our proposed life cycle includes cell division that occurs in the larger droplets, and sporulation for individual cells, enabling cell numbers to be replenished against loss.

In this work, we use the terms ''microbial life'' or ''microbes'' for microscopic life, without intending to imply that hypothetical Venusian microbes might in any way be taxonomically related to microbial life on Earth. We use the term ''spore'' to denote a cell in a dormant state of long-term metabolic inactivity, which is further resistant to (and protected from) environmental stresses.

We begin with a highlighted review on the very harsh and inhospitable conditions in the Venusian atmosphere and related, required assumptions for life to exist (Section 2). We next present our Venusian life cycle hypothesis (Section 3), which optimistically assumes that the challenges described in Section 2 can be met. We put the hypothesis in the context with Earth's aerial biosphere and other characteristics of the Venusian hypothesized aerial biosphere in Section 4. We conclude with a summary in Section 5.

## 2. Challenges and Assumptions for Life in the Venusian Clouds

The Venusian cloud decks are often described as a potentially habitable environment. The severe and unique environmental challenges, however, are often insufficiently explored. In this section, we review and emphasize the incredibly harsh conditions in the Venusian temperate clouds and cloud layer—far more extreme than any on Earth. Major assumptions must be made to envision life existing in such harsh conditions. After arguing why life must be confined to the inside cloud droplets, we review the challenges in order of severity.

### 2.1. Arguments for why microbial life ''outside the droplets'' is implausible

The requirement for a liquid environment is one of the general attributes of all life regardless of its biochemical makeup. If life's requirement for a liquid environment is universal, then on Venus the only conceivable stable habitat that meets this criterion is inside of cloud droplets.

Our main argument for why Venus atmosphere life outside liquid droplets is not possible is that free-floating life outside of droplets would rapidly desiccate (by net loss of liquid to the atmosphere) in the very dry atmosphere of Venus. Specifically, a free-floating cell will lose water until its internal water activity is the same as the vapor pressure of the atmosphere around it, which is more than an order of magnitude lower than the driest environments on Earth (Section 2.3). A non-desiccated cell will either have a water activity that is lower than that of the atmosphere around it, in which case water will condense onto it and it will then be in a droplet, or have a water activity that is higher than that of the atmosphere around it, and so will dry out. Only cells in a droplet will be stable. Dried out cells cannot actively divide.* To withstand the dry atmosphere of Venus, microbial life must use the protective environment of the inside of the droplet. In our proposed hypothesis, only one small part of the life cycle has desiccated cells in the form of spores that are dormant.

---

*We note that desiccation does not necessarily render active metabolism impossible. Several species of Earth microorganisms such as *Salmonella enterica*, *Listeria monocytogenes*, *Bradyrhizobium japonica*, and potentially others demonstrate time-variable transcriptional responses while desiccated, aimed at combating low water activity conditions with various osmoprotectants and membrane transporters (Cytryn *et al.*, 2007; Gruzdev *et al.*, 2011, 2012b, 2012a; Finn *et al.*, 2013b, 2013a).

On Earth, free-floating metabolically active cells outside of cloud water droplets are known to exist, but they are a small fraction of the overall aerial biomass. Free-floating cells are, however in danger of severe desiccation and death, if not deposited back on the surface within a few days. In addition, if such free-floating cells are swept up into the stratosphere, where the stellar UVC becomes the main sterilizing factor, they also die in a matter of days (Bryan *et al.*, 2019).

Published papers do not discuss the exact habitat for the hypothesized life in the Venusian clouds. One paper does link particles in the lower Venusian atmosphere (called Mode 3) to microbial life but only implicitly implies that life resides inside the Mode 3 droplets (Schulze-Makuch *et al.*, 2004). They specifically propose that the Mode 3 particles are microbes that are coated by elemental sulfur ($S_8$ or octathiocane). However, there is an inconsistency here whereby the hydrophobic nature of such pure octathiocane layers with no hydrophilic component would have prevented an efficient accumulation of liquid (both water or sulfuric acid) around the microbial cell, effectively exposing it to the outside atmosphere (Petrova, 2018). We note that if the shell is composed of both elemental sulfur and additional hydrophilic filaments, then active accumulation of liquids could, in principle, be possible.

### 2.2. The high concentrated sulfuric acid environment

The very high concentration of $H_2SO_4$ and extreme acidity is a unique challenge for life on Venus. Prior speculations about life on Venus often emphasize that terrestrial polyextremophiles can tolerate very low pH, high temperature, and low water activity environments with a "high concentration" of sulfuric acid. The implication is that the Venusian sulfuric acid cloud conditions have similarly low water activity and low pH, and hence are conducive to life (Morowitz and Sagan, 1967; Grinspoon, 1997; Cockell, 1999; Schulze-Makuch and Irwin, 2002, 2006; Schulze-Makuch *et al.*, 2004; Grinspoon and Bullock, 2007; Limaye *et al.*, 2018).

This is an incorrect implication. We cannot emphasize enough that the Venusian sulfuric acid clouds are much more acidic than even the most harshly acidic conditions found on Earth—the Dallol Geothermal Area, within the Danakil Depression in Northern Afar (Ethiopia) (Cavalazzi *et al.*, 2019). The Dallol acidic pools are high temperature (108°C), hypersaline (NaCl supersaturated), and anoxic hydrothermal sites containing up to 150 g/L of iron with the lowest environmental pH recorded to date (pH between −1.7 and 0) (Kotopoulou *et al.*, 2019).

Preliminary studies suggest that even in such harsh polyextreme hydrothermal conditions, life can survive and possibly even thrive; there are examples suggesting that the Dallol acidic pools are inhabited by very small (∼0.2 μm cell size) *Nanohaloarchaea* and other phyla of archaea (Belilla *et al.*, 2019; Gómez *et al.*, 2019). (Note that there is one Dallol pool where life has not yet been detected; instead of NaCl, the salts are composed of $Mg^{2+}$ and $Ca^{2+}$ ions.) The Dallol acidic pools, however, are not a model for Venusian clouds. The hyper-acidic environment of Dallol pools has water with sulfuric acid dissolved in them; they have the properties of water, even though they have a pH <0. In other words, the Dallol pools correspond to "only" ∼5% solution of $H_2SO_4$. By contrast, the Venusian cloud droplets are sulfuric acid with water dissolved in them.

The "pH" of Venusian clouds defined in a conventional way ($-\log_{10}[H^+]$) is meaningless because the conventional pH scale refers only to dilute aqueous solutions. The Hammett acidity value is a measure of acidity that naturally continues the pH scale up to concentrated acids such as sulfuric acid (Liler, 2012). The Hammett Acidity of 85% sulfuric acid is about −11.5 (Yates *et al.*, 1964). Acidity functions are on a log scale, so the clouds of Venus are $>10^{11}$ times as acidic as the Dallol geothermal area. This supports our statements that the Venusian cloud drops are an entirely different environment from any found naturally on Earth.

There is no Earth-based analogy of life adapting to or living in sulfuric acid concentrations as high as those in Venus cloud droplets. It is quite impossible for terrestrial metabolism to function in concentrated sulfuric acid where the majority of terrestrial biochemicals would be destroyed in seconds. There is extensive literature on the reactions of classes of molecules with concentrated sulfuric acid. Crucial biochemicals are unstable in sulfuric acid and include sugars (including nucleic acids, RNA, and DNA) (Krieble, 1935; Dische, 1949; Long and Paul, 1957), proteins (Reitz *et al.*, 1946; Habeeb, 1961), and other compounds such as lipids and complex carbohydrates (as shown, *e.g.*, by studies on dissolution of organic matter away from the outer shell of pollen grains; Moore *et al.*, 1991) and small-molecule metabolites of Earth's life core metabolism (Wiig, 1930; DeRight, 1934).

One might suggest that Venusian life is insulated from the sulfuric acid environment by elemental sulfur shells (Schulze-Makuch *et al.*, 2004). Elemental sulfur is present in the clouds of Venus (Taylor *et al.*, 2018). More importantly, elemental sulfur is not wetted by sulfuric acid (Petrova, 2018), so in principle it could provide chemical protection from the concentrated sulfuric acid environment. Cells coated with elemental sulfur shells would only adhere to the $H_2SO_4$ droplets rather than be entirely enveloped by them. By definition, hydrophobic particles cannot efficiently act as cloud condensation nuclei (CCN)—and thus this concept of protective sulfur shells is incompatible with our proposed life cycle, where the spores must act as CCN to become hydrated.

### 2.3. Very low water content

If Venusian life is water based, then the extremely dry Venusian atmosphere is a major challenge for life. Global water vapor mixing ratios in the Venusian atmosphere average around 40–200 ppm (Donahue and Hodges Jr, 1992; Barstow *et al.*, 2012). For context, the Venusian atmosphere's water content would be equivalent to a relative humidity of ∼0.07% at 298 K (Gmitro and Vermeulen, 1964); the driest air on Earth (excluding artificial laboratory conditions), the Atacama Desert at midday in summer, has a relative humidity of ∼2% (Cáceres *et al.*, 2007), so the environment of Venusian cloud droplets is literally ∼50 times as arid as the driest place on Earth.

Venusian water activity is much lower than any analogous habitat on Earth. On Earth, some species of filamentous fungi and yeasts are capable of growth at a water activity ($a_w$) as low as 0.61. Several haloarchaea are capable of growing in saturated NaCl environments with $a_w$ of 0.75 (Potts, 1994; Grant, 2004). If Venusian life uses a water-based solution for the cell interior, the life would have to employ special strategies aimed at water capture and retention.

The habitat inside liquid droplets has a higher water content than the atmosphere. The droplets are mostly $H_2SO_4$ (85% by volume, on average) with a much smaller component of liquid water (15%). The droplets in the clouds of Venus do have a varying composition from ~75% $H_2SO_4$ at high altitudes to ~110% (*i.e.*, $H_2SO_4$ with 10% $SO_3$) at the cloud base (Titov *et al.*, 2018) after (Hoffman *et al.*, 1980; James *et al.*, 1997). This is in equilibrium with the water vapor in the atmosphere. The activity of water in the droplets is very low, however, because the water is tightly bound to sulfuric acid molecules. The abundance of water as a function of altitude is poorly constrained by kinetic models, but it is in the low ppm range. A small fraction of water escapes the atmosphere as H or $H_2$ (Barabash *et al.*, 2007)—whether this is replenished or whether Venus is still slowly "drying out" is not known.

### 2.4. Nutrient scarcity

Permanently confined to an aerial biosphere, life must get all its nutrients from the atmosphere. We first review and comment on the availability of carbon, hydrogen, nitrogen, oxygen, phosphorus and sulfur (CHNOPS) biogenic elements before turning to the main nutrient scarcity, metals.

CHNOPS elements are all present in the Venusian atmosphere habitable layer as, for example, $CO_2$, $N_2$, $SO_2$, and $H_2O$, in modest amounts. Even phosphorus compounds are measured to be present in the Venusian atmosphere. The abundance of phosphorus in the Venus atmosphere and on the surface has only been measured by one of the Venera descent landers (although this X-ray-based detection determined the elemental P abundance it did not determine which P-bearing chemical species the phosphorus was in). In the altitude range of 52 and 47 km, the abundance of phosphorus appears to be on the same order as the abundance of sulfur (Vinogradov *et al.*, 1970; Surkov *et al.*, 1974; Andreichikov, 1987, 1998), as reviewed in Titov *et al.* (2018). Above 52 km no phosphorus was detected, and at 47 km the probe appeared to fail. Surface X-ray fluorescence of phosphorus species would have been masked by the much more common silicate.

There are examples on Earth of microbes that obtain all their C and N from the atmosphere. The chemolithoautotrophic and acidophilic bacterium *Acidithiobacillus ferrooxidans* can fix both $CO_2$ and $N_2$ from the atmosphere to build biomass (Valdés *et al.*, 2008; Quatrini and Johnson, 2019). *A. ferrooxidans* lives in extreme environments on Earth at low pH (1–2) and at moderately high temperatures of 50–60°C. Life in the Venusian aerial biosphere could use similar autotrophic strategies to that of *A. ferrooxidans* to obtain nutrients directly from atmospheric gases.

Venus is heavily depleted in hydrogen with respect to Earth. This is illustrated by the increased ratio of deuterium to hydrogen in the Venusian atmosphere. The ratio is ~100 times higher as compared with the rest of the solar system planets (Donahue *et al.*, 1982), which likely resulted from a catastrophic loss of water through atmospheric erosion by the solar wind. For comparison, on Earth $H_2$ reaches 0.55 ppm levels (Novelli *et al.*, 1999) whereas on Venus the amount of $H_2$ is much lower at 4 ppb (Krasnopolsky, 2010; Gruchola *et al.*, 2019). Such low abundance of hydrogen could result in hydrogen being a limiting nutrient for any aerial life that resides in the clouds of Venus. Production of hydrogen-rich, biochemically critical molecules, such as $CH_4$ or $NH_3$, might be very much more costly on Venus than on Earth due to the overall hydrogen scarcity.

Metals are scarce in the Venusian atmosphere, and the low metal abundance is likely to be a major growth-limiting factor. Metal ions are required for many biological functions (*e.g.*, more than a third of all proteins in terrestrial eukaryotes and bacteria require metal binding to function properly; Shi and Chance, 2011). Further, previous studies have speculated that Venusian life might use Fe/S redox chemistry as part of the biomass buildup and energy metabolism (Limaye *et al.*, 2018). It is conceivable that there are some amounts of mineral dust from the surface in the atmosphere of Venus and that a fraction of that dust could dissolve in sulfuric acid droplets, thus being accessible for life that resides inside the droplets.

Many metals are soluble in $H_2SO_4$, including many silicate ores and other metal salts (although glass and silica species are not). There is extensive literature dating many decades back about the study of metal salts in concentrated sulfuric acid, reviewed in part in Liler (2012). It may be worth acknowledging that although Earth-like biochemistry requires metals for biochemistry this may not be a universal requirement for all life. In principle, many biochemical functions of metals can be substituted by other specialized molecules built exclusively from biogenic CHNOPS elements (Hoehler *et al.*, 2020).

The scarcity of non-volatile nutrients, including metals, could, in principle, be mitigated by meteoritic delivery. The degree to which meteoritic delivery might be efficient is unknown. Although Venus is volcanically active, the planet lacks the global tectonic activity seen on Earth (*i.e.*, plate tectonics) (Byrne *et al.*, 2018). Only a few active volcanic sites have been detected on Venus' surface (Shalygin *et al.*, 2015; Treiman, 2017) and it is postulated that the overall volcanic flux is much lower than that on Earth (Mikhail and Heap, 2017). It is unknown whether the transport of dust from volcanic activity to the atmosphere is sufficient to be considered a viable source of non-volatile elements for the aerial biosphere in the clouds.

Life on Earth can inhabit environments where nutrients are extremely scarce. A recent example is the discovery of microbial ecologies in the lower oceanic crust (Li *et al.*, 2020). In such a nutrient-scarce environment, the key to life's survival includes: storage of materials as a source of elements in an event of extreme shortages; recycling and reusing of already acquired nutrients; and metabolic flexibility (*i.e.*, the ability to utilize multiple sources of carbon, nitrogen, energy, etc.). Similarly, life in Earth's aerial biosphere has developed a series of specific adaptations for efficient capture of limiting nutrients, including siderophore-mediated transition metal capture (see ahead to Fig. 4) (Amato *et al.*, 2019). It is conceivable that Venusian life, if it exists, has evolved similar solutions. If such solutions allowed for the efficient scavenging and recycling of metallic trace elements by Venusian life, then surface dust, meteoritic influx, or both could provide a sufficient source to compensate for losses.

### 2.5. Energy requirements and the assumption of photosynthesis

Living in the nutrient-depleted incredibly harsh conditions of the Venusian cloud decks—much more extreme than any conditions found on Earth—is likely incredibly energy intensive. For example, significant energy would be expended on such cellular processes as: active transport of substances through the cell membrane against the concentration gradient (*e.g.*, active import of water, if life on Venus is water based [see the example below]); acquisition of gaseous nutrients and fixation of elements into organic matter (*e.g.*, nitrogen fixation); biosynthesis of complex biochemicals that are crucial for nutrient storage, recycling, and retention (*e.g.*, siderophores for metal reuse and retention [see Section 2.4 above]); and, finally, motility if the microorganism needs to actively move within a droplet (*e.g.*, ciliary or flagellar movement).

We must assume that Venusian aerial microbial life is photosynthetic, so that energy capture is not a limiting factor. First, this is because there is little chemical energy potential within the atmosphere, and any redox disequilibria on the surface (the type of disequilibria exploited by terrestrial chemotrophs) are inaccessible to cloud-based life. Second, it is because of the abundance of solar energy on Venus and noting that the conversion of light energy to chemical energy through photosynthesis is the fundamental process in which life on Earth harnesses energy (the other being redox reactions). In fact, it has been speculated that the ''unknown UV absorber'' in the clouds of Venus, a chemical species of unknown identity, absorbing more than half of all UV that the planet receives could be a manifestation of a wide-spread energy-capture process by an aerial biosphere (Limaye *et al.*, 2018).

Photosynthesis need not be oxygenic, or even for carbon fixation. Life on Earth uses several strategies to photosynthetically fix carbon, with oxygenic photosynthesis being only one of the possibilities (reviewed in Seager *et al.*, 2012). Examples of anoxygenic photosynthesis include several types of sulfur-based photosynthesis, for example, used by green sulfur bacteria, or more exotic arsenic photosynthesis employed by *Ectothiorhodospira* bacterial communities from hot springs in Mono lake, California (Kulp *et al.*, 2008; Hoeft McCann *et al.*, 2017). Sulfur-based photosynthetic strategy was proposed as a potential means in which Venusian life could fix carbon (Schulze-Makuch *et al.*, 2004). We emphasize, however, that the use of light energy by life to run chemical reactions, generate and store energy is not limited to carbon fixation.

There are a number of terrestrial examples for utilization of sunlight for metabolic processes other than fixing carbon. (In principle, any chemical reaction or metabolic activity can be coupled to the light-capturing process.) Aphids, for example, have a layer of carotenoids under the cuticle that forms a sunlight-harvesting system coupled to the synthesis of adenosine triphosphate (ATP) (Valmalette *et al.*, 2012). Carotenoid pigments capture light energy and pass it on to the cellular machinery involved in energy production, a process completely independent from ''classical'' carbon fixation done in photosynthesis.

For a specific example of energy requirement and satisfaction, consider the assumption that Venusian life uses a water-based solution for the cell interior. In this case, the life would have to employ special strategies aimed at water capture and retention. The energy required to extract one mole of water from 85% sulfuric acid is $\sim 25$ kJ mol$^{-1}$ (depending on temperature). The solar energy available (solar flux between 380 and 740 nm) in the temperate cloud region depends on the altitude and varies from 636 J m$^{-2}$ s$^{-1}$ at 48 km to 952 J m$^{-2}$ s$^{-1}$ at 60 km (as calculated from models; P. Rimmer, private communication 2020). Considering a cell with a radius of 1 μm has a volume of about 4.2 fL, the energy needed to extract the corresponding $2.3 \times 10^{-13}$ moles of water is about $6 \times 10^{-9}$ J. The visible-wavelength solar energy available at 52 km is 760 J m$^{-2}$ s$^{-1}$, translating to about $9.5 \times 10^{-10}$ J for a 1 μm cell area. There is, therefore, plenty of light energy incident on the cell surface area. Energy from on the order of seconds to minutes of illumination for a microbe 1 μm in radius would be sufficient for life to extract sufficient water from its sulfuric acid environment to fill the organism.

UV radiation should not be a problem for any assumed Venusian life because of UV-protecting pigments, such as melanins. It is possible that melanins can harness high-energy electromagnetic radiation (a phenomenon called radiosynthesis) for useful metabolic activity. Some species of fungi (*e.g.*, *Cryptococcus neoformans*) can use this harsh ionizing radiation to promote growth (Dadachova *et al.*, 2007; Dighton *et al.*, 2008). The discovery of organisms that actively seek sources of highly ionizing radiation (they are radiotrophic) and preferentially grow in such environments (they were detected in space stations, Antarctic mountains, and in the nuclear reactor cooling water) opens the possibility that fungal melanins could have functions analogous to other energy harvesting pigments such as chlorophylls in algae and plants (Dadachova and Casadevall, 2008). So rather than solar UV being a problem for life at the top of the clouds, it could be a resource to be used. We note that the flux of ionizing radiation (galactic cosmic rays or extreme solar particle events) is only sterilizing high up in the atmosphere, and thus it is not a significant survival challenge for our putative Venusian aerial biosphere in the temperature cloud region (Dartnell *et al.*, 2015).

### 3. A Proposed Cycle for Venusian Aerial Microbial Life

We propose a life cycle for Venusian microbes (Fig. 1) that begins in the lower haze layer where desiccated spores reside in a dormant phase. The spores are transported upward by vertical mixing induced by gravity waves to a habitable layer of temperate conditions. Acting as CCN, the spores become encased in a liquid droplet (mostly $H_2SO_4$ with some $H_2O$) and germinate. During months aloft, the bacteria metabolize and divide. The cloud droplets, meanwhile, collide and grow to a size large enough that gravity forces them to settle downward. On the downward journey, triggered by changes in the environment (increasing temperatures and the concomitant evaporation of liquids), the bacteria sporulate, preserving themselves as desiccated spores. Once reaching the stable, long-lived stagnant lower haze layer ''depot,'' the spores remain dormant until the life cycle can begin again. In this section, we describe each step in more detail.

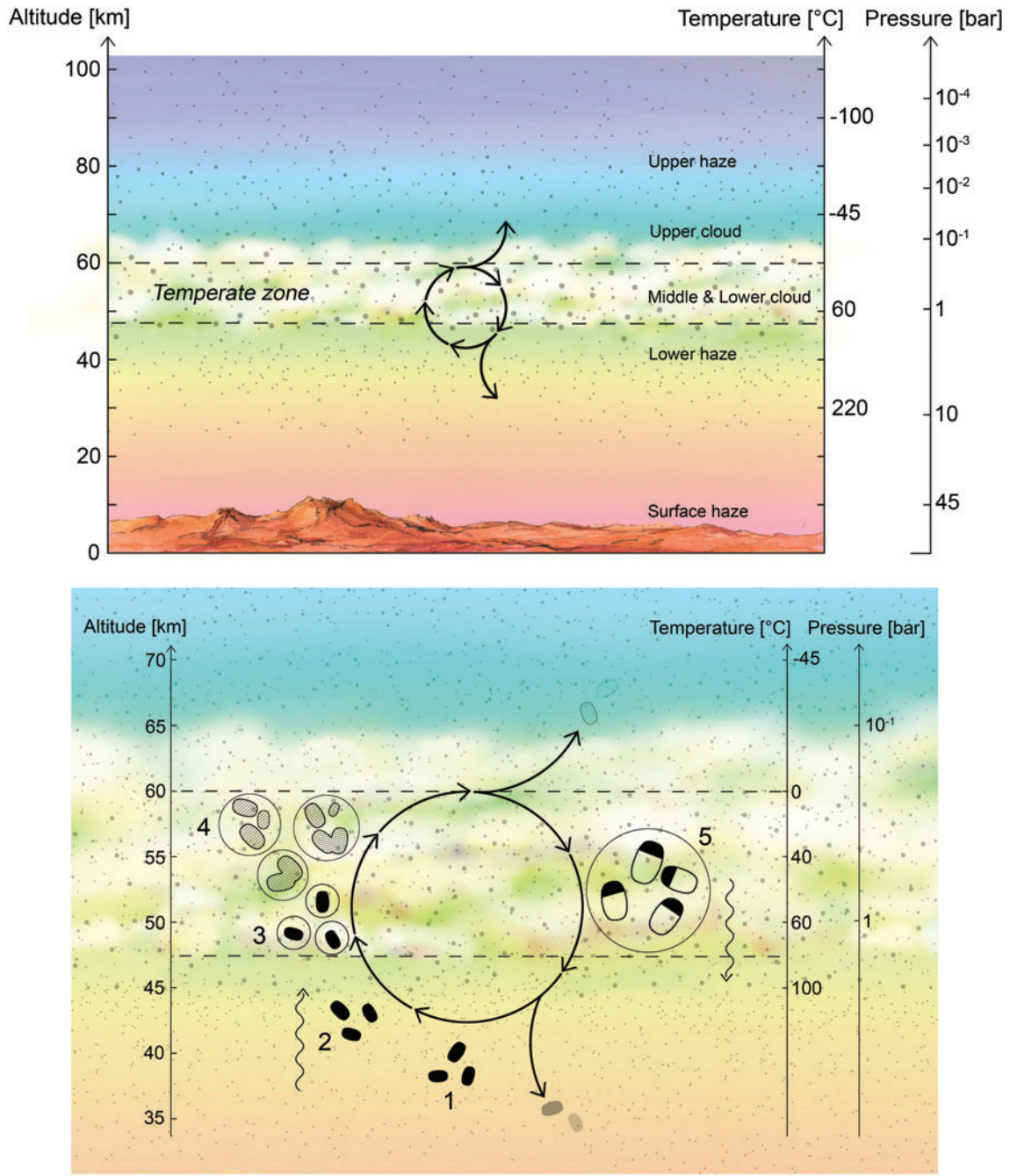

**FIG. 1.** Hypothetical life cycle of the Venusian microorganisms. Top panel: Cloud cover on Venus is permanent and continuous, with the middle and lower cloud layers at temperatures that are suitable for life. Bottom panel: Proposed life cycle. The numbers correspond to steps in the life cycle as described in the main text. (1) Desiccated spores (black blobs) persist in the lower haze. (2) Updraft of spores transports them up to the habitable layer. (3) Spores act as CCN, and once surrounded by liquid (with necessary chemicals dissolved) germinate and become metabolically active. (4) Metabolically active microbes (dashed blobs) grow and divide within liquid droplets (solid circles). The liquid droplets grow by coagulation. (5) The droplets reach a size large enough to gravitationally settle down out of the atmosphere; higher temperatures and droplet evaporation trigger cell division and sporulation. The spores are small enough to withstand further downward sedimentation, remaining suspended in the lower haze layer ''depot.'' CCN, cloud condensation nuclei. Color images are available online.

### 3.1. Step 1: Desiccated spores populate the lower Venus atmosphere haze layer, a depot of hibernating microbial life

Venus has a lower haze layer of relatively low mass and unknown composition (Titov *et al.*, 2018), such that desiccated bacterial spores could be a minor component. The lower haze layer resides below Venus' lowest clouds, at altitudes of 47.5 km down to 33 km, where temperatures range from 350 K to 460 K (Zasova *et al.*, 2006). Under pressures of around 10 bars (Taylor and Hunten, 2014), these temperatures are too high for water to exist in liquid form and are also high enough for sulfuric acid to thermally decompose (Knollenberg and Hunten, 1979; Knollenberg *et al.*, 1980; Krasnopolsky, 2013; Titov *et al.*, 2018). What is known about the haze layer is the particle size and number density (Fig. 2), derived from Pioneer Venus data reported in the canonical work by Knollenberg and Hunten (1980).

Desiccated spore sizes must be consistent with the small particle radii in the lower haze layer. This motivates the question of whether spores of that small size are large enough to contain all the required "cell machinery." Although the question of minimum cell size has only been considered for a hydrated free-living cell, it is likely similar for a spore. Cells must have sufficient space to accommodate metabolic machinery. Based on terrestrial life, a cell is unlikely to be much smaller than 0.2 μm in diameter based on the actual volume of genetic material, enzymatic complexes required for replication, transcription, and translation, in addition to a set of other proteins that contribute to basic physiological processes (Luef *et al.*, 2015; Chen *et al.*, 2018).

Although these minimal cell sizes are established for Earth's microbial life, it is unlikely that the hypothetical Venusian aerial life can be much smaller. Any complex metabolic activities would require complex biochemical machinery, which would require enough cell volume to function properly, even if the chemical basis of Venusian life is different than on Earth. Interestingly, the minimal recorded bacterial spore size on Earth is 0.25 μm diameter (Staley, 1999), with a more general range of 0.8 to 1.2 μm diameter (Krieg and Holt, 1984; Ricca and Cutting, 2003). For comparison, mean particle radii in Venus' lower haze layer generally range from 0.2 to 0.5 μm, but with radii as large as 2 μm at the top of the lower haze layer (Knollenberg and Hunten, 1980) (Fig. 2). Therefore, the particle size distribution in the Venusian atmosphere lower haze layer is compatible with the known range of cell and spore sizes of Earth's microorganisms.

We refer to the lower haze layer as a depot due to its relative stagnation compared with the rest of the Venus atmosphere, allowing desiccated bacteria to persist for a prolonged period.

For example, the sedimentation timescale, $\tau_{\text{sed}}$, of lower haze particles is >100 years (Fig. 3), as defined by

$$\tau_{\text{sed}} = \frac{H}{v}, \quad (1)$$

where $H$ is the scale height (on the order of 5 km; Seiff *et al.*, 1985) and $v$ is the sedimentation velocity (Fig. 3). In addition, the lower haze layer is stable against convection, meaning that vertical transport by convection does not occur. The stable stratification inference is based on *in situ* lapse rate measurements by the Pioneer Venus probes (Schubert *et al.*, 1980). Therefore, there is no convective overturning to transport haze particles upward, or downward to the hot, deep atmosphere (Schubert *et al.*, 1980).

Venus has a Hadley cell flow that moves upward at the equator and downward at the poles, but not enough is understood to know what altitude particles trapped in the flow may be lofted to or deposited in. In addition, the location of the returning branch is not known and thus may not impact the lower haze (Sánchez-Lavega *et al.*, 2017).

### 3.2. Step 2: Desiccated spores in the lower Venus atmosphere haze layer travel up to the lower clouds by mixing via gravity waves, followed by convective entrainment

The spores in the lower haze layer must be transported upward to continue their life cycle, but the relative stagnation of the lower haze layer, as described in Section 3.1, creates a challenge. One possible solution is the action of gravity waves, which appear to be present in the lower haze layer due to the layer's static stability. Although gravity waves can only lead to the net transport of energy and momentum and not matter, they can compress atmosphere layers as they travel and contribute to atmospheric mixing.

The gravity waves could be launched by convective plumes arising in the adjacent (50–55 and ∼18–28 km) convective regions (Schubert *et al.*, 1980; Seiff, 1983; Baker *et al.*, 2000a, 2000b; Lefèvre *et al.*, 2018). Radio occultation observations found wave-like features in intensity and temperature (Woo and Ishimaru, 1981; Hinson and Jenkins, 1995) that are attributed to gravity waves. Upward and downward vertical winds likely associated with gravity waves balance out, with vertical velocities of ∼1 m s$^{-1}$, as measured directly by the Venera landing probes 9 and 10 at the lower haze layer altitudes (and below) with anemometers (Kerzhanovich and Marov, 1983). Other entry probes indirectly measured windspeeds from Doppler spectroscopy, radio Doppler tracking, or acoustic measurements (Sánchez-Lavega *et al.*, 2017).

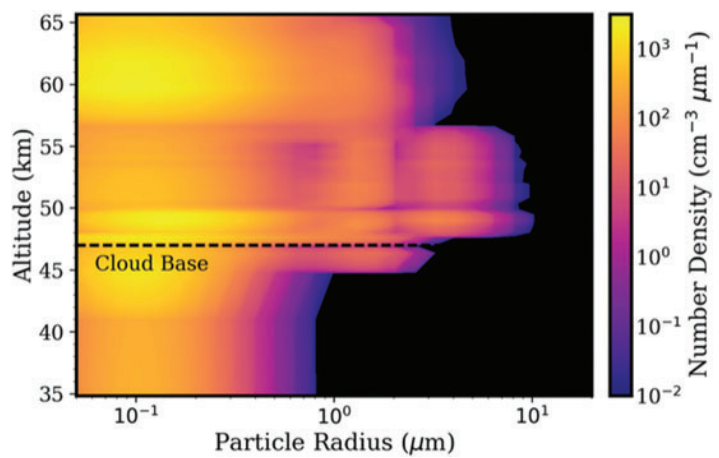

**FIG. 2.** Cloud and haze particle size distribution as a function of altitude. The contour colors show the number density of particles. Particles in the lower haze layer (33–47.5 km) are predominantly smaller than 1 μm (relevant to Step 1 of the proposed life cycle). Particles in the lower and middle clouds (47.5–56.5 km) can extend in radius to well more than 1 μm (relevant to Step 3 of the proposed life cycle). Data from Knollenberg and Hunten (1980), Table 4. Color images are available online.

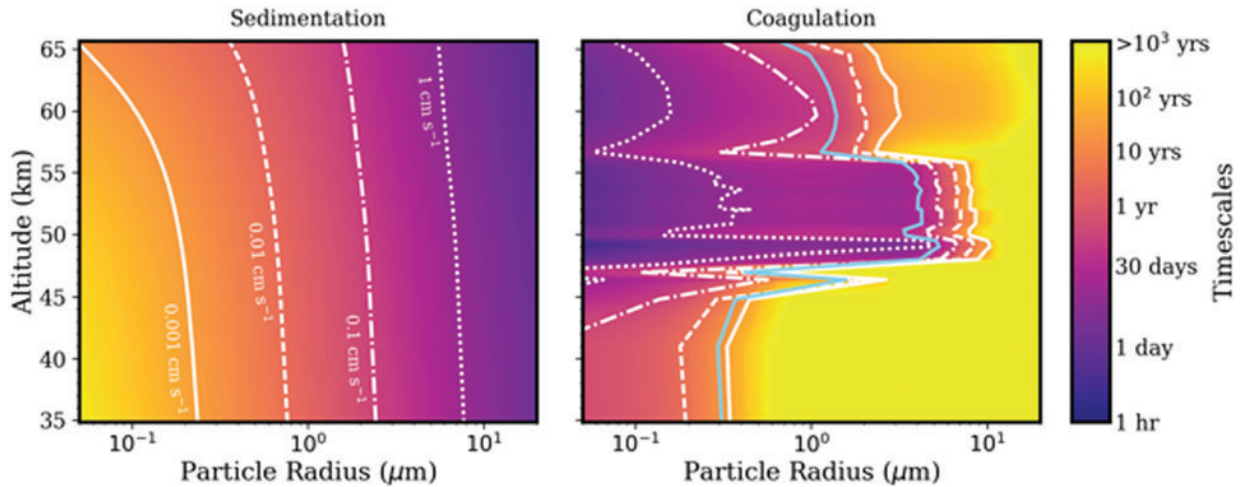

**FIG. 3.** Sedimentation velocities (left) and growth by coagulation timescales (right) as a function of particle radius and altitude. The contour colors show timescales, whereas the white curves show sedimentation velocities on the left, and sedimentation velocities needed to quench coagulation on the right. The blue curve on the right shows where the sedimentation timescale equals the coagulation timescale. Small particles in the lower haze layer (35–47.5 km) are long lived against sedimentation with timescales of tens to thousands of years for particles smaller than 1 μm (relevant to Step 1). Particles in the lower and middle cloud layers (47.5–56.5 km) coagulate on the timescale of hours to months, and particles on order 1 to a few microns in size may be stable against sedimentation for months to years (relevant to Step 4). Sedimentation and coagulation timescales are calculated by using data from Knollenberg and Hunten (1980), table 4, and the Venus International Reference Atmosphere (Seiff *et al.*, 1985). Color images are available online.

We hypothesize that gravity waves mix some of the spores in the lower haze into the lower clouds, where they can participate in cloud formation. We parameterize the mixing by using eddy diffusion, which approximates turbulent transport in planetary atmospheres. Particles, if small enough, can follow the gas in eddy diffusion (Lilly, 1973). The eddy diffusion coefficient ($K_{zz}$) in the Venus lower haze layer has been inferred from radio occultations and chemical models to be ∼10$^4$ cm$^2$ s$^{-1}$ (Woo and Ishimaru, 1981; Krasnopolsky, 2007). The time, $t$, for vertical diffusion is (Gifford, 1968; Jacob, 1999),

$$t = \frac{d^2}{2K_{zz}} \quad (2)$$

where $d$ is vertical distance. Given length scales of 1 to 10 km, the diffusion time is ≤1 year. This timescale is significantly shorter than the sedimentation timescale [Eq. (1)], such that the desiccated spores can be mixed upward before they sediment downward out of the lower haze layer. However, the sedimentation timescale can decrease if the particles were to grow larger.

Gravity waves operate in both vertical directions, meaning that some desiccated spores will be lost to lower, hotter layers. Thus, the lower haze layer may be more of a ''leaky'' depot for desiccated spores.

In the lower haze, where temperatures are too warm for condensational growth, coagulation may play an important role. The coagulation timescale, $\tau^i_{coag}(s)$, for particles with radius $i$ is defined as

$$\tau^i_{coag} = \left( \sum_{j > i} K_{ij} n_j \right)^{-1}, \quad (3)$$

where $n_j$ is the number density of particles with radius $j$ and $K_{ij}$ is the coagulation kernel between particles with radius $i$ and particles with radius $j$. Here, we have computed the coagulation kernel for the conditions of the lower haze by using the standard coagulation kernel equations that include Brownian motion, convection, and gravitational collection (Pruppacher and Klett, 1978), taking into account the temperature, viscosity, density, and molecular weight of the atmosphere (Seiff *et al.*, 1985). The kernel also requires the haze particle radius, mass, and sedimentation velocity, yielding kernel values for every pair of colliding particle radii. We compute the timescale by using the coagulation kernel and the observed particle number density as a function of particle radius (from Knollenberg *et al.*, 1980), yielding a minimum coagulation timescale of ∼1 year, assuming every collision leads to sticking (Dominik and Tielens, 1997). The measured particle sizes in the lower haze layer, however, are small, indicating that growth by coagulation is not a common process, such that mixing may be what ultimately controls the lower haze particle distribution. We note that if cloud particles are charged (from ionizing charges, Michael *et al.*, 2009), such charge effects on particles might reduce coagulation substantially.

In summary, although the dynamics of the lower haze layer are highly uncertain, upward (and downward) transport of haze particles is likely accomplished through mixing via gravity waves. Once transported upward to the bottom of the lower cloud layer, particles may continue to efficiently move upward into the clouds by convective entrainment.

### 3.3. Step 3: The desiccated spores act as CCN and once surrounded by a liquid droplet germinate to a metabolically active life-form

We have argued in Section 2.1 that active microbial life must live inside a droplet, so the spores must be only part of the life cycle. This section describes how the spores become engulfed in cloud droplets.

Once transported to the Venus lower cloud layer, the spores must act as CCN. CCN are "cloud seeds," a small solid surface needed for vapor to condense. Unlike the relatively high temperature at lower haze altitudes, the lower cloud layer has temperatures where liquid and vapor sulfuric acid and water can coexist. Sulfuric acid ($H_2SO_4$) vapor is produced photochemically from $SO_3$ and water vapor, with more $H_2SO_4$ being produced at higher altitudes because of the higher flux of UV radiation.

There is a precedent for spores acting as CCN. Aerial bacteria on Earth act as CCN for ice nucleation (Morris *et al.*, 2004; Creamean *et al.*, 2013). The cells have ice-nucleating proteins to catalyze the formation of ice nuclei (IN) from liquid (Pandey *et al.*, 2016). Although aerial bacteria are suspected to act as CCN for vapor condensation to liquid based on lab experiments (Bauer *et al.*, 2003), there is not yet any definitive *in situ* evidence.

The Venusian spores are likely to have a hydrophilic and hygroscopic exterior so they can attract and absorb both sulfuric acid vapor and water vapor. We note that terrestrial bacteria capture water by using hygroscopic biosurfactant polymers. Many microbial polysaccharides and amphipathic lipopeptides, such as syringafactin, from *Pseudomonas syringae* have highly hygroscopic properties and are instrumental in reducing the water stress of microorganisms (Burch *et al.*, 2014). Once self-encased within a $H_2SO_4$ and $H_2O$ cloud droplet, the Venusian spore would become solvated, eventually leading to the revival and full physiological activation of the cell. On Earth, spores reactivate when the environmental conditions become favorable for active metabolism and growth. The revival of spores occurs on rehydration and subsequent swelling and the reactivation of metabolic activity (Sella *et al.*, 2014).

We envision that the Venusian spores only constitute a small fraction of CCN, not large enough to affect prior CCN concepts and calculations. The CCN have previously been suggested to be photochemically generated polysulfur compounds (Toon *et al.*, 1982, 1984; Imamura and Hashimoto, 2001). This is a problematic suggestion, because reduced polysulfur solids are likely to be hydrophobic and therefore not able to act as a CCN (Young, 1983; Petrova, 2018). Both meteoritic dust and $FeCl_3$ upwelled from the surface have also been proposed as CCN (Turco *et al.*, 1983; Krasnopolsky, 1985, 2017; Gao *et al.*, 2014). Once nucleation occurs, the initial growth through condensation is rapid, with a timescale on the order of seconds (James *et al.*, 1997).

### 3.4. Step 4: The cellular life-form lives in the droplet for months to years, depending on the path of the droplet—during this time, the droplet grows by coagulation

Once in the Venus lower clouds the droplets grow and circulate around the atmosphere. The particles will collide, and each collision for liquid particles results in coagulation, leading to further droplet growth. Coagulation timescales [Eq. (3)] are on the order of days to months in the temperate cloud layers (Fig. 3) and particles may grow to sizes greater than 1 µm. Note that at the same time, zonal flow or Hadley cell motions can carry the droplets around the planet with a timescale of days to months (Schubert *et al.*, 1980), but this is not relevant to particle growth.

Once the droplet reaches a large enough size ($\sim 1\,\mu m$), the cell residing inside the droplet has room to grow and divide. The timescales on which droplets persist in the habitable layer depend on particle size and altitude and are controlled by droplet growth by coagulation and sedimentation due to gravity (Fig. 3). The range can be hours to months to years.

The critical motivating fact for our life cycle description is that droplets will continue to grow by coagulation until the sedimentation timescale becomes shorter than the coagulation timescale, and the particles fall rapidly into deeper, hotter layers of the atmosphere. (Falling much faster than the diffusive transport.) A key question is then: Do the microbes have enough time to metabolize and divide before their droplet home falls to an altitude where they must form spores to survive? The answer is yes. For example, a 3 µm-radius particle in the Venusian lower clouds persists for about 6 months, which should be more than enough time for cellular life to germinate from the spore, metabolize, grow, and divide within the same droplet.

Another valid question is whether the droplets in Venus' temperate cloud decks provide enough habitable space. We argue yes. The most numerous droplets in the Venusian middle and lower clouds are 2 µm in diameter (1 µm in radius) (Fig. 2). As an example, majority of free-living soil bacteria and archaea have a cell diameter smaller than 0.5 µm, with some as small as 0.2 µm diameter (Hahn, 2004; Portillo *et al.*, 2013). The smallest free-living cell size on Earth is $\sim 0.2\,\mu m$ diameter (Luef *et al.*, 2015), leaving a 2 µm droplet diameter (1 µm droplet radius) enough room to accommodate a few cells. Cloud droplets larger than the common 2 µm diameter could contain relatively large microbial communities (droplets on the order of 10 µm diameter or larger could host dozens of cells). The large volume could enhance promotion of cell division.

The small size of some species of microorganisms does not mean that they are biochemically simple, or primitive from the evolutionary standpoint. For example, the smallest known free-living photosynthetic organism is the prokaryote *Prochlorococcus*, which is 0.5 to 0.7 µm in diameter (Chisholm *et al.*, 1988; Biller *et al.*, 2015).

As an interesting aside, *Prochlorococcus* is an example of life adapting to elemental scarcity. Living in a phosphorus-poor environment, *Prochlorococcus* substitutes sulfur for phosphorus in many of its biochemicals. In more detail, *Prochlorococcus* synthesizes sulfoquinovosyldiacylglycerol (SQDG), a membrane lipid that contains organic sulfate group ($-SO_3^-$) and sugar instead of phosphate, and uses it almost exclusively as a membrane lipid (>99% of membrane lipids of *Prochlorococcus* contain sulfate instead of phosphate groups) (Van Mooy *et al.*, 2006). Such substitution of sulfur for phosphorus is likely a crucial adaptation, and a reason for the tremendous evolutionary success of picocyanobacteria in nutrient scarce oligotrophic environments.

Although phosphorus is not believed to be a limiting element in Venus' clouds (Section 2.4), one can envision similar, analogous approaches for other elements as well.

To close out this part of the life cycle, there are two additional peripheral issues. Some particles might updraft up out of the habitable part of the clouds (above 60 km) by being trapped in the Hadley cell flow or by upward diffusion. We note, however, that surviving freezing temperatures in the Venus upper clouds is less challenging than surviving the scorching heat of the lower altitudes. Synthesis of a variety of cryoprotectants is a common strategy employed by life on Earth, including aerial bacteria in Earth's atmosphere, to mitigate extreme cold (Amato *et al.*, 2019). Life might even adapt to higher altitude layers where the temperature falls significantly below 0°C. Similar to the temperature, the concentration of sulfuric acid in cloud droplets changes drastically with altitude. As sulfuric acid remains liquid over a wide-range of temperatures and pressures (Gable *et al.*, 1950; Ohtake, 1993), even cloud droplets residing in the higher cloud decks could remain liquid, where the temperatures and pressures are low (*e.g.*, at 60 km the temperature falls below 0°C and pressure falls below 0.5 bar). Life could adapt to temperatures as low as −20°C. In fact, surviving in low temperatures is likely much easier to achieve than in high temperature regimes. Earth bacteria routinely survive freezing. Moreover, some obligatory psychrophiles such as *Colwellia psychrerythraea* 34H are commonly found in sea ice with liquid brine temperatures as low as −35°C (Maykut and Untersteiner, 1986), have active motile behavior in temperatures as low as −10°C (Junge *et al.*, 2003), and actively reproduce in temperatures as low as −5°C (Huston, 2004). Any particles at high altitudes will eventually return to the habitable layer via the Hadley cell flow, eddy diffusion, or sedimentation. However, at high altitudes high UV irradiation could be destructive if the microbes do not have a strongly UV protective layer (c.f. Section 2.5).

A second peripheral issue is that of droplet fragmentation. If fragmentation could occur, many droplets could avoid growing to sizes with large corresponding sedimentation velocities. More importantly, fragmentation would increase the microbial population in the habitable layer by creating new droplet habitats, some already populated with microbes from prior cell division. By populating the aerial biosphere by fragmentation, the lower haze depot would not be needed.

However, fragmentation of droplets should not occur in Venus' atmosphere. Fragmentation is quantitatively captured by the Weber number, $We$, which represents the ratio of disruptive hydrodynamic forces to the stabilizing surface tension forces (Pilch and Erdman, 1987; Testik and Gebremichael, 2010), $We = \frac{\rho v^2 d}{\sigma}$. Here, $\rho$ is the density of the flow field (*i.e.*, atmosphere), $v$ is the initial relative velocity between the flow field and the drop, $d$ is the initial diameter of the drop, and $\sigma$ is the surface tension of the drop. There is a critical Weber number below which droplet fragmentation does not occur, $We < 12$, (as long as the droplet's viscous forces are less than the combined inertial and surface tension forces; Pilch and Erdman, 1987). For the Venus atmosphere habitable cloud layer: The atmosphere density is 0.4 to 1.5 kg m$^{-3}$; most particles are 1 to a few microns in size falling at maximum 1 cm s$^{-1}$ (Figs. 2 and 3); the surface tension for a mixture of 10% water and 90% H$_2$SO$_4$ by volume at a range of temperatures (273–323 K) is 55 (mN m$^{-1}$) (Myhre *et al.*, 1998), resulting in $We \ll 1$, well below the Weber critical number for fragmentation.

### 3.5. Step 5: The bacteria settle down out of the clouds as the droplet reaches a maximum size that can stay aloft against gravity—the decreasing liquid activity triggers cell division and sporulation

Venusian life, trapped living inside of liquid droplets, must adapt to the eventual downward droplet migration to the lower, hotter parts of the Venusian atmosphere. Temperatures during the downward fall increase, and this would result in a gradual loss of liquid activity. In other words, the droplets begin to evaporate, and conditions become inhospitable for physiologically active cells. We hypothesize that in response to degrading conditions (high temperature and low liquid activity), microbes begin metabolic preparations for sporulation and deposition of desiccated spore cells. Desiccated to very small sizes, the spores remain in the lower haze, dormant until upward eddy diffusion caused by mixing induced by gravity waves brings them back to the habitable layer.

Sporulation is the ability of various organisms on Earth to form small, desiccated, and metabolically inactive cells, called endospores, or more generally spores (Setlow, 2006). Sporulation is an adaptation to detrimental environmental conditions that aims at preserving the genetic material of the cell when the surrounding environment is inhospitable or lethal for metabolically active cells. The triggers for sporulation on Earth are stress factors, including dehydration, nutrient limitation, and high cell density (Setlow, 2006; Hutchison *et al.*, 2016). Sporulation allows microorganisms to wait out unfavorable conditions and persist in a dormant state until environmental conditions become favorable again. On Earth, bacterial endospores are superbly resistant to high and low temperatures, high and low pressures, desiccation, radiation, and toxic chemicals (Nicholson *et al.*, 2000). We hypothesize that an analogous process to Earth bacterial sporulation could happen as an essential step during the life cycle of Venusian life. Although sporulation of Earth organisms is often an occasional event triggered as a last resort (Hutchison *et al.*, 2016), as a response to unpredictable environmental conditions, on Venus periodical sporulation events are crucial for microbes' long-term survival.

The final step in our propsed Venusian life cycle is when the dessicated spores settle out into the lower haze layer. As the spores lose liquid and become less massive, their downward settling times slow further. The individual cells must have a coating that prevents cells clumping together as they dry out, such that each individual spore is a single microbial cell. This critical step is the only way to sustain stable numbers of cells in the cloud decks, as it is the only step in the life cycle where the number of distinct atmospheric particles that contain cells can increase (the number of cells increases during the growth phase in the clouds, but the number of particles containing cells remains the same because the cells all remain in one droplet). For example, fungal, bryophyte, and other spores produced in a dense clump in a sporangium (or equivalent) scatter on release because of the properties of the spore-forming body (Sundberg and Rydin, 1998). Each spore remains dormant until mixing brings it back to the beginning of the life cycle (Section 3.2).

Spores could remain viable in the Venusian atmosphere lower haze layer for long periods, based on analogy with Earth life. On Earth, some bacterial spores can survive in extremely harsh conditions for many thousands of years (Nicholson *et al.*, 2000; Paul *et al.*, 2019). The generation times of some subseafloor sediment bacteria has been estimated to be thousands of centuries (Parkes *et al.*, 2000). More relevant are spores that are viable but have been dormant for thousands of years (Christner *et al.*, 2000; Aouizerat *et al.*, 2019). The currently held, although highly controversial, record for the longest time spent in suspended animation belongs to the spores of the species of *Bacillus* sealed in a salt crystal that formed 250 Mya ago (Vreeland *et al.*, 2000). Although so far only one such extremely old example is known and the study is controversial as the age of the isolated bacterium is frequently questioned (Maughan *et al.*, 2002), it suggests that Earth-based bacterial spores may be capable of survival in metabolically inactive states for at least several million years (Cano and Borucki, 1995; Meng *et al.*, 2015). Bacteria and archaea are not the only organisms capable of surviving prolonged periods of harsh environmental conditions. Some complex, multicellular organisms (*e.g.*, tardigrades) can remain in suspended animation for several years (Guidetti and JoÈnsson, 2002), as can the dormant eggs of killifish *Nothobranchius*, which can survive as dried eggs for more than three years and be hatched successfully (Cellerino *et al.*, 2016).

### 4. Discussion

With our hypothesized life cycle articulated, we now turn to a discussion for further context.

#### 4.1. Earth's aerial biosphere

Clouds on Earth harbor a diverse species of microbial life, including bacteria, archaea, eukaryotes, and viruses (Amato *et al.*, 2017, 2019). On Earth, most microbes reside inside cloud liquid water droplets but some are free-floating in the atmosphere. In the past, scientists were skeptical of microbial survival in Earth's clouds, because of: UV-fluxes, low temperatures, severe desiccation, and the transient and fragmented status of cloud cover that could lead to a sudden loss of habitat. Within the past decade, however, this harsh view of Earth clouds as a habitat for life has improved. Modern molecular biological techniques such as metagenomics and metatranscriptomics have given new insights into microbial diversity and metabolic functioning, especially for microbes residing inside cloud water droplets (Amato *et al.*, 2017, 2019) (Fig. 4).

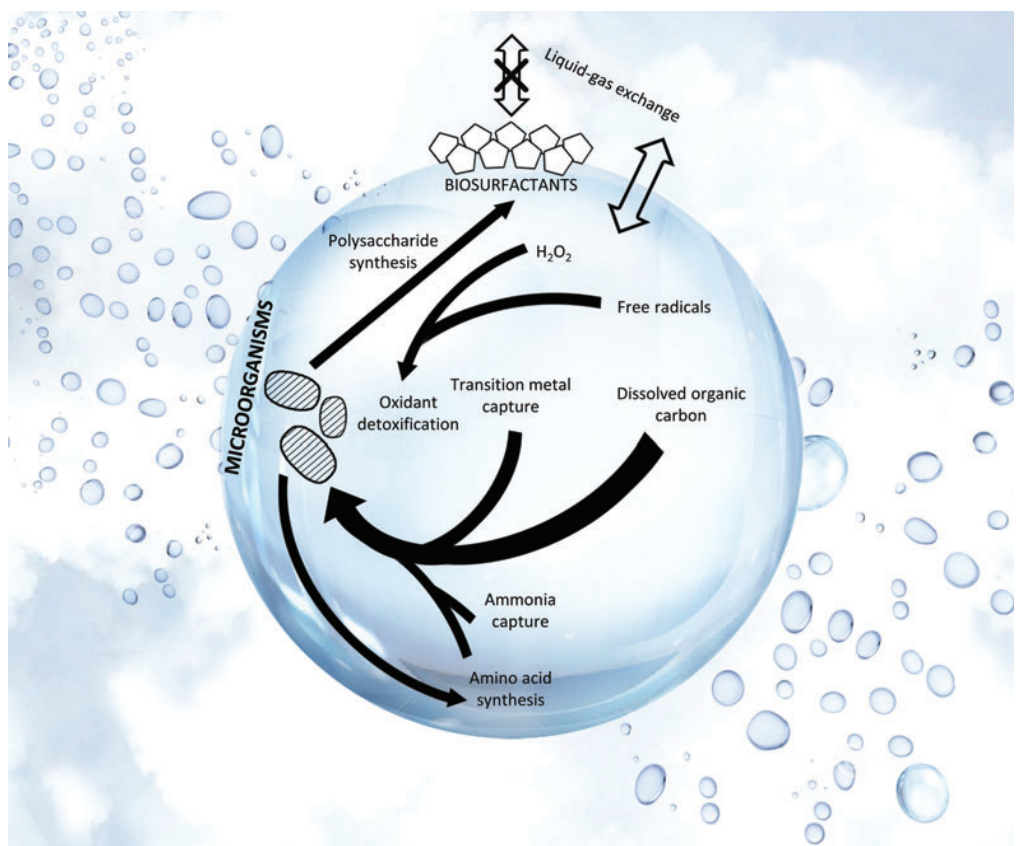

**FIG. 4.** Major metabolic processes for life residing in Earth cloud liquid water droplets. Single-celled microorganisms are shown by the dashed blobs. Selected key adaptations for microbes to survive inside the droplet are depicted by black arrows. Thicker arrows represent more important pathways. The processes were identified from the metatranscriptomics studies of Earth's aerial cloud biosphere (Amato *et al.*, 2019). The droplet size has a diameter on order of 10 µm. The background is meant to be illustrative of cloud makeup. Figure adapted from Amato *et al.* (2019). Color images are available online.

The transport of microbes from Earth's surface up into the clouds is now known to be a common phenomenon (Vaïtilingom *et al.*, 2012; Amato *et al.*, 2017; Bryan *et al.*, 2019) (Fig. 5). Earth's cloud ''aerial biosphere'' is believed to serve as a temporary refuge during long distance transportation across the planet, rather than a permanent habitat for microbial life.

Microbes are eventually deposited to the surface by precipitation (Vaïtilingom *et al.*, 2012). On Earth, bacterial aerosols in the troposphere remain aloft on average for

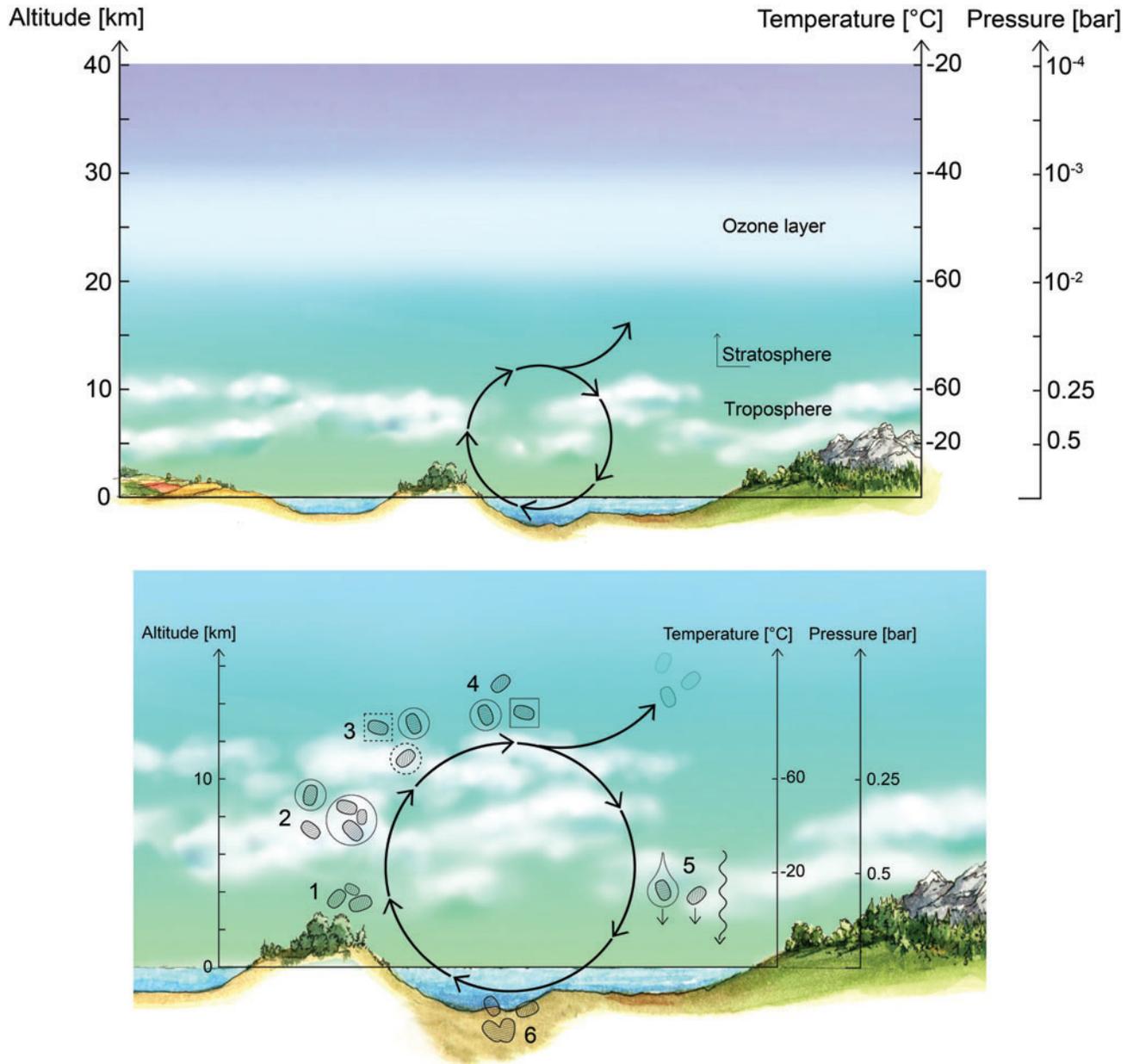

**FIG. 5.** The life cycle of Earth's aerial biosphere is intimately connected with the habitable surface. Top panel: The cloud cover on Earth is transient and fragmented and is, therefore, not a permanent habitat for Earth's aerial biosphere. (This is in contrast to Venus where the cloud cover is permanent and continuous.) Bottom panel: Life cycle of Earth's aerial biosphere. (1) Updraft of metabolically active microorganisms (dashed blobs) from the surface. (2) Microbial cells are metabolically active both within water cloud droplets (solid circles) and in the free-floating form. (3) Cells likely act as CCN (dashed circle) and promote IN (dashed square) in the atmosphere, promoting droplet formation. (4) Metabolically active cells transiently persist in the atmosphere, are transported over long distances until (5) deposition onto the surface by precipitation or downdraft. (6) On colonization of the new surface habitat, active cell division commences. There is, as of yet, no evidence for cell division in the clouds. Note that Earth's microbial aerial biosphere is metabolically active at every step of the life cycle, and survival is not limited to microorganisms capable of sporulation (Bryan *et al.*, 2019). We note that some fraction of free-floating cells can be transported high into the stratosphere (∼38 km) where, if they are not brought back down within a couple of days, they will die from severe desiccation and high-UV exposure (semi-transparent dashed blobs). IN, ice nucleation. Color images are available online.

3 to 7 days (Burrows *et al.*, 2009), which is enough time for microbes to be transported over long distances, across whole continents and oceans (Barberán *et al.*, 2015; Griffin *et al.*, 2017; Šantl-Temkiv *et al.*, 2018).

Earth bacteria swept up from the surface have been postulated to act as CCN of water clouds (Sattler *et al.*, 2001; Bauer *et al.*, 2003) or trigger ice nucleation. However, this has not been proven to occur outside the lab; calcium carbonate shells (coccoliths) are as yet the only known biological source that can act as CCN (Trainic *et al.*, 2018).

Earth's clouds are a challenging ecological niche for permanent habitation because of their transient and fragmented nature (in contrast to Venus' permanent and continuous cloud cover). In addition to the fact that Earth's surface is habitable, there is no evolutionary pressure exerted on the microbial biosphere on Earth to adopt a life cycle permanently sustained in the clouds. Rather, the evolutionary selection has likely been focused on temporary survival in the clouds (including cloud-specific complex metabolic functions) in anticipation for the eventual deposition on the habitable surface of the planet. For clouds to be a permanent habitat, active cell division would have to occur in the clouds. Metatranscriptomics studies on Earth's aerial biosphere have not identified transcripts related to active DNA replication and cell division, suggesting that DNA replication and cell division are not performed *in situ* by metabolically active cells in the cloud biosphere. Therefore, Earth's microorganisms are using the clouds to migrate to new habitats, not to stay in the clouds and reproduce. This implies that the cloud droplets are only transiently inhabited by microbial life and that the aerial biosphere is intimately tied to the habitable surface.

Although there is no direct evidence of active cell division available *in situ* in cloud droplets, there is ample evidence for a surprisingly physiologically active and diverse metabolism of microbes in cloud droplets (Amato *et al.*, 2019). Diverse physiological and biochemical strategies of microbes have been identified that seem to be direct, specific adaptations to cloud droplet environment. Those include protection against oxidants, osmotic pressure variations, the synthesis of cryoprotectants to fight extreme cold, or production of metal ion scavengers and biosurfactants (Amato *et al.*, 2017, 2019) (Fig. 4). Earth cloud droplets could, therefore, greatly extend the layer of the biosphere that is at least transiently inhabited by metabolically active life on Earth. Further, clouds are definitively not the upper limit of the biosphere, as there is evidence for free-floating viable microbial isolates in the stratosphere up to 38 km (Bryan *et al.*, 2019). We emphasize, however, that, contrary to cells inhabiting cloud droplets, such free-floating microbial life will only survive a few days due to severe desiccation and high UV exposure (Bryan *et al.*, 2019) (see Section 2.1 for discussion of the specific challenges for free-floating microbial life in the atmosphere).

The characteristics that enable certain microorganisms to retain viability in clouds are likely attributed to the selective advantage of being able to survive long-distance transport to new surface habitats that is provided by temporary cloud droplet colonization (Fig. 5). The habitability of the Earth surface likely squashed any potentially significant evolutionary advantages that might be gained from permanent cloud colonization.

In contrast to Earth, the lack of habitable surface on Venus would force hypothetical microbes to live in the clouds permanently. Therefore, the natural selection pressure on Venus would be directed toward evolutionary strategies that allow life to colonize clouds permanently and not just transiently as it is on Earth. This forms the basis for our life cycle hypothesis: that the entire life cycle of Venusian life, including genetic material replication and cell division, must occur in the only temperate region of Venus—the clouds.

### 4.2. Active mechanisms for remaining aloft?

Throughout the article, we have assumed that the passive mechanisms of air movement drive our proposed life cycle, enabling microbes to remain aloft in the Venusian temperate zone for part of their life cycle. Active mechanisms might evolve but are beyond the scope of this article. Active mechanisms could perpetuate life in the clouds of Venus in four ways. First, life could be part of the life cycle proposed here, actively propelling the desiccated spores upward over many kilometers.

Second, life could maintain falling droplets in the cloud layer (*i.e.*, propel falling droplets upward). Third, life could expel living cells or spores from falling droplets in the cloud layer, which the vertical air movements discussed earlier could then mix throughout the clouds to colonize other, smaller droplets. On Earth some microscopic fungi and bacteria are capable of pushing fruiting bodies through the surface tension of water droplets and into the air phase before releasing spores through the air to new habitats, in the process successfully breaking surface tension at the water–air interface (Talbot, 1999; Elliot and Talbot, 2004). This is a relatively complex, active mechanism, and it only works over cm to m distance scales. Lastly, fourth, at least in principle, life could be droplet-independent macroscopic life with the ability to self-transport.

### 4.3. On biomass and fluxes in the Venusian lower haze layer

We are far from being able to construct a plausible ecosystem with biomasses and fluxes, due to lack of information. One estimate, focusing on the abundance of particles (Mode 2 and Mode 3) in the Venus lower and middle clouds, is that the biomass could be 0.1 to 100 mg m$^{-3}$, which is comparable to Earth's aerial 44 mg m$^{-3}$ (Limaye *et al.*, 2018). In this work, we can, for a minimum, use approximate estimates to at least show that our proposed life cycle has "legs to stand on." Say 1/100,000 of the lower haze layer mass is desiccated biomass. The mass density of the entire haze layer is estimated to be in the range of 0.1–2 mg m$^{-3}$ (Titov *et al.*, 2018) from Gnedykh *et al.* (1987), leading to a lower range of mass in the lower haze layer on the order of 500 million metric tons. Taking 1/100,000 gives 5,000 tons for the desiccated biomass in the lower haze layer. Assuming desiccated particles have a mass of 0.4 pg (Bratbak and Dundas, 1984; Perry *et al.*, 2002) brings us to about $10^{22}$ number of spores. The total number of cells in the Venusian aerial biosphere could be 10 times or greater than the number in the lower haze, or $10^{23}$ total particles, as the proposed life cycle includes cell division while inside droplets aloft.

For the context, Earth is estimated to have more than $10^{30}$ free-living single-celled microbes (Whitman *et al.*, 1998;

Bar-On *et al.*, 2018), with a rough estimate of $10^{24}$ total cells in Earth's aerial biosphere (Bryan *et al.*, 2019). The Venusian atmosphere is likely to support only the tiniest fraction of Earth's total biomass, because the only habitable part of Venus is the temperate cloud layer, with the habitable volume likely further limited by available nutrients. Earth, in contrast, is essentially habitable in its entirety, including the atmosphere, surface, and subsurface with all those habitats interconnected.

For our proposed life cycle to be viable, microbes within droplets must reproduce in large enough numbers (*e.g.*, each droplet falling down could have dozens of organisms in it) to balance the spores lost from the haze layer. We imagine the Venus atmosphere as three boxes: the clouds, the lower haze, and the deep hot atmosphere that acts as a sink to the microbes. In the clouds, the total mass of microbes can increase by uptake of nutrients from the droplet or ambient atmosphere, leading to cell division, such that the downward flux of cells from the clouds can be larger than the upward flux from the lower haze. The excess downward flux of cells, however, must balance the downward flux from the lower haze into the deep atmosphere, where the microbes are lost forever to fatally high temperatures. If we conservatively imagine that 10 times more spores are lost to the lower atmosphere than get transported back up (and further accounting for some spores that just die off), each cell would have to divide, say four times (reaching 16 cells per large droplet). This is very reasonable, considering that in Earth's anaerobic conditions, cell division can be as short as the order of hours, not the months or years that the droplets remain in the cloud layer.

The Venusian atmosphere dynamics, especially in the lower haze layer, are not well enough understood to have any certainty to work out mass or flux balance. All we can say is that upward mixing by gravity waves is a way to move small particulate biomass upward, and would have to approximately equal the planet-wide but slow sedimentation of biomass trapped inside droplets. Vertical transport and fluxes, including Hadley cell transport at the equator and poles, definitely require further study.

## 5. Summary

Life in the Venusian clouds has long since been a popular if speculative topic. The clouds decks of Venus themselves are often described as conductive to life. We reassessed this notion and reviewed the severe and unique environmental challenges that life in the aerial biosphere of Venus would have to overcome. The challenges include: an extremely acidic environment, far more so than any known environment on Earth; very low water content; and nutrient scarcity.

We also highlighted the assumption that life would have to reside inside protective cloud droplets (sulfuric acid mixed with water), and that any life would have to be photosynthetic to have enough energy for a variety of cellular processes.

The main new point of this work is to present a life cycle concept. Assuming that life must reside inside cloud droplets, we resolve the subsequent conundrum of gravitationally settling droplets reaching hotter, uninhabitable regions by proposing a Venusian life cycle where a critical step is microbes drying out to become spores on reaching the relatively stagnant lower haze layer, which we call a leaky "depot." The dried out spores would reside there until some of them can be transported back up to the temperate, habitable cloud layers, where they would act as CCN to promote cloud formation, becoming enveloped in cloud droplets to continue the life cycle.

### Acknowledgments

The authors thank Joanna Petkowska-Hankel for the preparation of Figs. 1, 4, and 5. They also thank Daniel Koll for useful discussions.

### Author Disclosure Statement

No competing financial interests exist.

### Funding Information

The authors thank the Change Happens Foundation, the Heising-Simons Foundation, and the MIT Professor Amar G. Bose Research Grant Program for funding. P.G. acknowledges support from the 51 Pegasi b Fellowship funded by the Heising-Simons Foundation.

A PROPOSED VENUSIAN LIFE CYCLE 15

Address correspondence to:
*Sara Seager*
*Department of Earth, Atmospheric,*
*and Planetary Sciences*
*Massachusetts Institute of Technology*
*54–1718, 77 Massachusetts Avenue*
*Cambridge, MA 02139*
*USA*

*E-mail:* seager@mit.edu

**Abbreviations Used**

CCN = cloud condensation nuclei
CHNOPS = carbon, hydrogen, nitrogen, oxygen, phosphorus and sulfur biogenic elements